\documentclass[11pt,a4paper]{article}

\usepackage{hyperref}
\usepackage{amsmath, amsthm, amssymb}
\usepackage{graphicx}
\usepackage[square,sort,comma,numbers]{natbib}

\title{Comment on ``Aharonov-Bohm Phase is Locally Generated Like All Other Quantum Phases"}

\begin{document}

\author{Shan Gao
\\Research Center for Philosophy of Science and Technology, 
\\ Shanxi University, Taiyuan 030006, P. R. China
\\ E-mail:  \href{mailto:gaoshan2017@sxu.edu.cn}{gaoshan2017@sxu.edu.cn}.}

\maketitle

\begin{abstract}
Marletto and Vedral [Phys. Rev. Lett. 125, 040401 (2020)] propose that the Aharonov-Bohm (AB) phase is locally mediated by entanglement between a charged particle and the quantized electromagnetic field, asserting gauge independence for non-closed paths. In this Comment, we critically analyze their model and demonstrate that the AB phase arises from the interaction with the vector potential \(\mathbf{A}\), not from entanglement, which is a byproduct of the quantum electrodynamics (QED) framework. We show that their field-based energy formulation, intended to reflect local electromagnetic interactions, is mathematically flawed due to an incorrect prefactor and yields \( +q \mathbf{v} \cdot \mathbf{A}_{\mathbf{s}} \) in the Coulomb gauge, conflicting with QED’s \( -q \mathbf{v} \cdot \mathbf{A}_{\mathbf{s}} \). This equivalence to \( q \mathbf{v} \cdot \mathbf{A}_{\mathbf{s}} \) holds only approximately in the Coulomb gauge under static conditions, failing for time-dependent fields and other gauges, undermining their claim of a gauge-independent local mechanism. Furthermore, we confirm that the AB phase is gauge-dependent for non-closed paths, contradicting their assertion.  Our analysis reaffirms the conventional explanation in the semi-classical picture, where the AB phase is driven by the vector potential \(\mathbf{A}\), with entanglement playing no causal role in its generation. 
\end{abstract}

\section{Introduction}

The Aharonov-Bohm (AB) effect \citep{aharonov1959} demonstrates that a charged particle’s wavefunction acquires a phase due to the vector potential \(\mathbf{A}\), even in regions where electromagnetic fields vanish. For a closed path, the phase is:

\begin{equation}
\phi_{AB} = \frac{q}{\hbar} \oint \mathbf{A} \cdot d\mathbf{l} = \frac{q \Phi}{\hbar},
\end{equation}
where \(q\) is the particle’s charge and \(\Phi\) is the magnetic flux. This phenomenon, highlighting the physical significance of gauge potentials, raises questions about its local versus non-local nature, as the phase depends on the enclosed flux, suggesting a non-local interaction. Marletto and Vedral \citep{marletto2020} propose a quantum field theory (QFT) model asserting that the AB phase is locally mediated by entanglement between the charged particle and the quantized electromagnetic field, claiming gauge independence for non-closed paths and detectability via local measurements.

In this Comment, we critically examine their model using quantum electrodynamics (QED). We demonstrate that the AB phase arises from the coupling between the charge’s current and the solenoid’s current via the photon propagator, with the vector potential \(\mathbf{A}\) serving as an effective description. 
Their field-based energy, intended to reflect local electromagnetic interactions, is flawed due to an incorrect prefactor and yields \( +q \mathbf{v} \cdot \mathbf{A}_{\mathbf{s}} \) in the Coulomb gauge, contradicting QED’s \( -q \mathbf{v} \cdot \mathbf{A}_{\mathbf{s}} \). This equivalence holds only approximately under static conditions in the Coulomb gauge, failing for time-dependent fields or other gauges, undermining their gauge-independent local mediation claim. We also confirm that the AB phase is gauge-dependent for non-closed paths, contradicting their assertion. 
Additionally, we argue that entanglement, while present in the QED framework, is not the primary driver of the phase, which is fundamentally governed by the interaction with the vector potential \(\mathbf{A}\). Our analysis seeks to clarify the mechanisms underlying the AB effect and reaffirm its conventional interpretation within QED.

\section{Marletto and Vedral's Model}

Marletto and Vedral \citep{marletto2020} model a charged particle (charge \(q\), mass \(m\)) in a superposition of paths around a solenoid, using qubits for the charge (\(|0\rangle_C, |1\rangle_C\)) and solenoid (\(|0\rangle_S, |1\rangle_S\)). The electromagnetic field is quantized with photon operators \(a_k, a_k^\dagger\). Their Hamiltonian in the Coulomb gauge (after the signs of the interaction terms are corrected) is: 

\begin{eqnarray}
H_{AB} &=& E_C q_z^{(C)} + E_S q_z^{(S)} + \int d^3 \mathbf{k} \hbar \omega_k a_k^\dagger a_k \nonumber \\ 
&-& \int d^3 \mathbf{k} g_k \frac{q}{m} \mathbf{p} \cdot \mathbf{u}_k \left( a_k e^{i \mathbf{k} \cdot \mathbf{r}_{\mathbf{c}}} + a_k^\dagger e^{-i \mathbf{k} \cdot \mathbf{r}_{\mathbf{c}}} \right) q_z^{(C)} \nonumber \\
&-& \int d^3 \mathbf{k} \int d^3 \mathbf{x} g_k \mathbf{j} \cdot \mathbf{u}_k \left( a_k e^{i \mathbf{k} \cdot \mathbf{x}} + a_k^\dagger e^{-i \mathbf{k} \cdot \mathbf{x}} \right) q_z^{(S)},\label{Hint}
\end{eqnarray}
where $E_C$ and $E_S$ are the free energies of the charge and the solenoid, $q_z^{(C)}$ and $q_z^{(S)}$ are Pauli operators for the charge qubit and the solenoid qubit, $\omega_k$ and ${\bf k}$ represent the photon frequency and wavenumber of the mode respectively, \(\mathbf{p}\) is the charge’s momentum operator, \(\mathbf{r}_{\mathbf{c}}\) is its position, and \(\mathbf{j}(\mathbf{x} - \mathbf{r}_{\mathbf{s}})\) is the solenoid’s current density centered at \(\mathbf{r}_{\mathbf{s}}\). The coupling constant is \(g_k = \sqrt{\frac{\hbar}{2 \epsilon_0 \omega_k V}}\), and \(\mathbf{u}_k\) is the photon polarization vector satisfying \(\mathbf{k} \cdot \mathbf{u}_k = 0\). The vector potential is:

\begin{equation}
\mathbf{A}(\mathbf{x}) = \int d^3 \mathbf{k} g_k \mathbf{u}_k \left( a_k e^{i \mathbf{k} \cdot \mathbf{x}} + a_k^\dagger e^{-i \mathbf{k} \cdot \mathbf{x}} \right).
\end{equation}

They compute the phase from the transition amplitude:

\begin{equation}
\langle 1 |_C \langle 0 |_F \langle 1 |_S \exp \left( -\frac{i}{\hbar} H_{AB} \tau \right) | 1 \rangle_C | 0 \rangle_F | 1 \rangle_S = \exp \left\{ -i \left( \xi + \phi(\mathbf{r}_{\mathbf{c}}, \mathbf{r}_{\mathbf{s}}) \right)  \right\}.
\end{equation}
where $\tau$ is set to $1$. The phase \(\phi(\mathbf{r}_{\mathbf{c}}, \mathbf{r}_{\mathbf{s}})\) is computed from the interaction Hamiltonian using the second-order term of the time-ordered exponential, which can be shown to be 
\begin{equation}
\phi = -\frac{q}{m} \frac{\mu_0}{4 \pi\hbar} \int d^3 \mathbf{x} \frac{\mathbf{p} \cdot \mathbf{j}(\mathbf{x} - \mathbf{r}_{\mathbf{s}})}{|\mathbf{r}_{\mathbf{c}} - \mathbf{x}|}.
\end{equation}
The negative sign in the phase is missed in Marletto and Vedral's paper \citep{marletto2020}. The corresponding interaction energy is:

\begin{equation}
\mathcal{E} = \phi \hbar = -\frac{q}{m} \mathbf{p} \cdot \left( \frac{\mu_0}{4 \pi} \int d^3 \mathbf{x} \frac{\mathbf{j}(\mathbf{x} - \mathbf{r}_{\mathbf{s}})}{|\mathbf{r}_{\mathbf{c}} - \mathbf{x}|} \right).
\end{equation}
Using \(\mathbf{p} = m \mathbf{v}\), this becomes:

\begin{equation}
\mathcal{E} = -q \mathbf{v} \cdot \left( \frac{\mu_0}{4 \pi} \int d^3 \mathbf{x} \frac{\mathbf{j}(\mathbf{x} - \mathbf{r}_{\mathbf{s}})}{|\mathbf{r}_{\mathbf{c}} - \mathbf{x}|} \right).
\end{equation}
The vector potential in the Coulomb gauge is defined as 

\begin{equation}
\mathbf{A}(\mathbf{r}_{\mathbf{c}}) = \frac{\mu_0}{4 \pi} \int d^3 \mathbf{x} \frac{\mathbf{j}(\mathbf{x} - \mathbf{r}_{\mathbf{s}})}{|\mathbf{r}_{\mathbf{c}} - \mathbf{x}|},
\end{equation}
so:

\begin{equation}
\mathcal{E} = -q \mathbf{v} \cdot \mathbf{A}.\label{e1}
\end{equation}
For a solenoid with flux \(\Phi = B_0 \pi a^2\), \(\mathbf{A} = \frac{\Phi}{2 \pi (x^2 + y^2)} (-y, x, 0)\), and with \(\mathbf{v} = v \hat{y}\):

\begin{equation}
\mathcal{E} = -\frac{q v \Phi x}{2 \pi (x^2 + y^2)} = -\frac{q v B_0 a^2 x}{2 (x^2 + y^2)},\label{e2}
\end{equation}
yielding the accumulating phase during a time interval:

\begin{equation}
\phi_{AB} = \int \frac{\mathcal{E}}{\hbar} \, dt = -\frac{q}{\hbar} \int \mathbf{A} \cdot \mathbf{v} \, dt =  -\frac{q}{\hbar} \int \mathbf{A} \cdot dl, \label{eq:phase}
\end{equation}
which matches the semi-classical AB phase.

The above derivation in the Coulomb gauge is a specific case. It can be shown that the interaction energy calculated from the second-order term of the transition amplitude in QED with a quantized electromagnetic field for the magnetic AB effect always takes the form $-q \mathbf{v} \cdot \mathbf{A}$, where $\mathbf{A}$ is the classical vector potential in the chosen gauge. This holds for all gauges because the second-order amplitude, governed by the photon propagator and conserved currents, consistently yields the effective vector potential interaction. 

Marletto and Vedral start with a correct QED Hamiltonian in the Coulomb gauge (Eq. \ref{Hint}), modeling the interaction between a charged particle, solenoid, and quantized field. Based on the model, one can derive the standard interaction energy \(\mathcal{E} = -q \mathbf{v} \cdot \mathbf{A}\) (Eq. \ref{e1}), and correctly obtain the AB phase (Eq. \ref{eq:phase}). 
However, they propose a field-based energy, \(\mathcal{E}_{\text{field}} = \frac{1}{2} \int \left( \frac{\mathbf{B}_0 \cdot \mathbf{B}_{\mathbf{c}}}{\mu_0} + \epsilon_0 \mathbf{E}_{\mathbf{s}} \cdot \mathbf{E}_{\mathbf{c}} \right) d^3\mathbf{r}\) (Eq. \ref{eq:field_energy}), claiming it mediates the phase locally via entanglement and is gauge-independent. Their critical error, as we will argue below, lies in assuming \(\mathcal{E}_{\text{field}}\) is valid in all gauges, and in attributing the phase to entanglement rather than \(\mathbf{A}\). 

\section{Critique of the Claims}

\subsection{Marletto and Vedral's Field-Based Energy Proposal}
\label{subsec:field_energy}

Marletto and Vedral propose an alternative formulation of the interaction energy as a field overlap integral:
\begin{equation}
\mathcal{E}_{\text{field}} = \frac{1}{2} \int_V \left( \frac{\mathbf{B}_0 \cdot \mathbf{B}_{\mathbf{c}}}{\mu_0} + \epsilon_0 \mathbf{E}_{\mathbf{s}} \cdot \mathbf{E}_{\mathbf{c}} \right) d^3\mathbf{r},
\label{eq:field_energy}
\end{equation}
where $\mathbf{B}_0$ is the solenoid's magnetic field, $\mathbf{B}_{\mathbf{c}}$ is the field generated by the moving charge, and \(\mathbf{E}_{\mathbf{s}}\) and \(\mathbf{E}_{\mathbf{c}}\) are the electric fields of the solenoid and the particle, respectively. They claim this energy reflects a local electromagnetic (EM) field interaction that mediates the AB phase, independent of the vector potential \(\mathbf{A}\). We demonstrate that this formulation is flawed, both mathematically and physically, and that the AB phase is fundamentally driven by \(\mathbf{A}\), not local EM fields, which vanish along the charge’s path. In particular, we provide a detailed proof that their field-based energy is equivalent to the interaction energy \(q \mathbf{v} \cdot \mathbf{A}\) only in the Coulomb gauge, explicitly highlighting where this gauge is used.

\subsubsection{Mathematical Critique of the Field-Based Energy}

The \(\frac{1}{2}\) prefactor in Eq.~(\ref{eq:field_energy}) is inappropriate for the interaction energy between two distinct sources (the solenoid and the charged particle). In electromagnetic theory, the interaction energy between two systems with magnetic fields \(\mathbf{B}_0\) (solenoid) and \(\mathbf{B}_{\mathbf{c}}\) (charged particle) and electric fields \(\mathbf{E}_{\mathbf{s}}\) and \(\mathbf{E}_{\mathbf{c}}\) is given by:
\begin{equation}
\mathcal{E}_{\text{field}} = \frac{1}{\mu_0} \int_V \mathbf{B}_0 \cdot \mathbf{B}_{\mathbf{c}} d^3 \mathbf{r} + \epsilon_0 \int_V \mathbf{E}_{\mathbf{s}} \cdot \mathbf{E}_{\mathbf{c}} d^3 \mathbf{r}.
\label{eq:correct_field_energy}
\end{equation}
The \(\frac{1}{2}\) prefactor, typically used for the total field energy of a single system, leads to an underestimation of the interaction energy by a factor of 2, making Marletto and Vedral’s formulation quantitatively incorrect. 

\subsubsection{Equivalence to \(q \mathbf{v} \cdot \mathbf{A}\) in the Coulomb Gauge}

It can be demonstrated that the proposed field-based energy is equivalent to the interaction energy \(q \mathbf{v} \cdot \mathbf{A}\) in the Coulomb gauge. 

For a solenoid with steady current, the electric field \(\mathbf{E}_{\mathbf{s}} = 0\), simplifying the interaction energy to:
\begin{equation}
\mathcal{E}_{\text{field}} = \frac{1}{\mu_0} \int_V \mathbf{B}_0 \cdot \mathbf{B}_{\mathbf{c}} d^3 \mathbf{r},
\label{eq:simplified_field_energy}
\end{equation}
where \(\mathbf{B}_0 = B_0 \hat{z}\) for \(r < a\) (inside the solenoid, with radius \(a\)) and \(\mathbf{B}_0 = 0\) for \(r > a\), and \(\mathbf{B}_{\mathbf{c}}\) is the magnetic field produced by the charged particle (charge \(q\), velocity \(\mathbf{v} = v \hat{y}\)) at position \(\mathbf{r}_{\mathbf{c}} = (x, y, z)\).

To prove that \(\mathcal{E}_{\text{field}}\) is equivalent to the  interaction energy \(q \mathbf{v} \cdot \mathbf{A}\) in the Coulomb gauge, we evaluate Eq.~(\ref{eq:simplified_field_energy}) using a vector identity and explicitly note the use of the Coulomb gauge. The magnetic field \(\mathbf{B}_{\mathbf{c}} = \nabla \times \mathbf{A}_{\mathbf{c}}\), where \(\mathbf{A}_{\mathbf{c}}\) is the vector potential of the charged particle. We use the vector identity:
\begin{equation}
\mathbf{B}_0 \cdot \mathbf{B}_{\mathbf{c}} = \mathbf{B}_0 \cdot (\nabla \times \mathbf{A}_{\mathbf{c}}) = \nabla \cdot (\mathbf{A}_{\mathbf{c}} \times \mathbf{B}_0) + \mu_0 \mathbf{A}_{\mathbf{c}} \cdot \mathbf{j}_{\mathbf{s}},
\label{eq:vector_identity}
\end{equation}
where \(\mathbf{j}_{\mathbf{s}}\) is the solenoid’s current density, and we have used \(\nabla \times \mathbf{B}_0 = \mu_0 \mathbf{j}_{\mathbf{s}}\) inside the solenoid (since \(\mathbf{E}_{\mathbf{s}} = 0\)). Integrating over the volume \(V\):
\begin{equation}
\int_V \mathbf{B}_0 \cdot \mathbf{B}_{\mathbf{c}} d^3 \mathbf{r} = \int_{\partial V} (\mathbf{A}_{\mathbf{c}} \times \mathbf{B}_0) \cdot d\mathbf{S} + \mu_0 \int_V \mathbf{A}_{\mathbf{c}} \cdot \mathbf{j}_{\mathbf{s}} d^3 \mathbf{r},
\label{eq:integral_split}
\end{equation}
where the first term is a surface integral over the boundary \(\partial V\). Since \(\mathbf{B}_0 = 0\) outside the solenoid (\(r > a\)), the surface integral vanishes (since \(V\) encloses the solenoid and extends to a region where \(\mathbf{B}_0 = 0\)). Thus:
\begin{equation}
\mathcal{E}_{\text{field}} = \frac{1}{\mu_0} \int_V \mathbf{B}_0 \cdot \mathbf{B}_{\mathbf{c}} d^3 \mathbf{r} = \int_V \mathbf{A}_{\mathbf{c}} \cdot \mathbf{j}_{\mathbf{s}} d^3 \mathbf{r}.
\label{eq:field_energy_reduced}
\end{equation}

Now we need the vector potential \(\mathbf{A}_{\mathbf{c}}\) of the charged particle. In the Coulomb gauge (\(\nabla \cdot \mathbf{A}_{\mathbf{c}} = 0\)), the vector potential for a point charge \(q\) moving with velocity \(\mathbf{v}\) at position \(\mathbf{r}_{\mathbf{c}}\) is approximately (in the non-relativistic limit):
\begin{equation}
\mathbf{A}_{\mathbf{c}}(\mathbf{r}) = \frac{\mu_0 q \mathbf{v}}{4 \pi |\mathbf{r} - \mathbf{r}_{\mathbf{c}}|},
\label{eq:particle_vector_potential}
\end{equation}
where \(\mathbf{r}\) is the field point. 
Substituting into Eq.~(\ref{eq:field_energy_reduced}):
\begin{equation}
\mathcal{E}_{\text{field}} = \int_V \left( \frac{\mu_0 q \mathbf{v}}{4 \pi |\mathbf{r} - \mathbf{r}_{\mathbf{c}}|} \right) \cdot \mathbf{j}_{\mathbf{s}}(\mathbf{r}) d^3 \mathbf{r}.
\label{eq:field_energy_integral}
\end{equation}
The solenoid’s vector potential at the charge’s position \(\mathbf{r}_{\mathbf{c}}\) is:
\begin{equation}
\mathbf{A}_{\mathbf{s}}(\mathbf{r}_{\mathbf{c}}) = \frac{\mu_0}{4 \pi} \int \frac{\mathbf{j}_{\mathbf{s}}(\mathbf{r})}{|\mathbf{r} - \mathbf{r}_{\mathbf{c}}|} d^3 \mathbf{r},
\label{eq:solenoid_vector_potential}
\end{equation}
which is also defined in the Coulomb gauge (\(\nabla \cdot \mathbf{A}_{\mathbf{s}} = 0\)). Comparing this with Eq.~(\ref{eq:field_energy_integral}), we see:
\begin{equation}
\mathcal{E}_{\text{field}} = q \mathbf{v} \cdot \left( \frac{\mu_0}{4 \pi} \int \frac{\mathbf{j}_{\mathbf{s}}(\mathbf{r})}{|\mathbf{r} - \mathbf{r}_{\mathbf{c}}|} d^3 \mathbf{r} \right) = q \mathbf{v} \cdot \mathbf{A}_{\mathbf{s}}(\mathbf{r}_{\mathbf{c}}).
\label{eq:field_energy_equivalence}
\end{equation}
Thus, in the Coulomb gauge, the field-based energy reduces to the  interaction energy:
\begin{equation}
\mathcal{E}_{\text{field}} = q \mathbf{v} \cdot \mathbf{A}_{\mathbf{s}},
\label{eq:equivalence}
\end{equation}
matching Eq.~(\ref{e1}) in the original model, confirming the equivalence in the Coulomb gauge.

However, in non-Coulomb gauges (\(\mathbf{A}_{\mathbf{s}}' = \mathbf{A}_{\mathbf{s}} + \nabla \chi\)) with a time-independent $\chi$ (or without scalar potential), \(\mathbf{A}_{\mathbf{s}}\) includes additional terms, and thus the equivalence \(\mathcal{E}_{\text{field}} = q \mathbf{v} \cdot \mathbf{A}_{\mathbf{s}}\) does not hold. 
This means that the equivalence between the field-based energy and the  interaction energy $q \mathbf{v} \cdot \mathbf{A}_{\mathbf{s}}$ is gauge-specific, and they are not equivalent for non-Coulomb gauges. 

\subsubsection{Discrepancy Between Field-Based and QED Interaction Energies in the Coulomb Gauge}

Now we demonstrate that this field-based energy, \(\mathcal{E}_{\text{field}} = \epsilon_0 \int \mathbf{E}_{\mathbf{s}} \cdot \mathbf{E}_{\mathbf{c}} \, d^3 \mathbf{r} + \frac{1}{\mu_0} \int \mathbf{B}_0 \cdot \mathbf{B}_{\mathbf{c}} \, d^3 \mathbf{r}\), is not equal to QED’s interaction energy, \(\mathcal{E}_{\text{QED}}\), even in the Coulomb gauge, due to a sign error in the magnetic term and its failure to account for time-dependent fields accurately.

In QED, the interaction Hamiltonian for a charged particle in the Coulomb gauge (\(\nabla \cdot \mathbf{A}_{\mathbf{s}} = 0\)) is derived from minimal coupling:

\begin{equation}
H_{\text{int}} = -\frac{q}{m} \mathbf{p} \cdot \mathbf{A}_{\mathbf{s}} + q \phi_{\mathbf{s}},
\end{equation}
where \(\phi_{\mathbf{s}}\) is the scalar potential of the source. The interaction energy is the expectation value: 
\begin{equation}
\mathcal{E}_{\text{QED}} = \langle H_{\text{int}} \rangle = q \phi_{\mathbf{s}}(\mathbf{r}_{\mathbf{c}}) - q \mathbf{v} \cdot \mathbf{A}_{\mathbf{s}}(\mathbf{r}_{\mathbf{c}}),
\end{equation}
since \(\langle \mathbf{p} \rangle = m \mathbf{v}\). This energy includes both the electrostatic potential energy \(q \phi_{\mathbf{s}}\) and the magnetic interaction \(-q \mathbf{v} \cdot \mathbf{A}_{\mathbf{s}}\), valid for both static and time-dependent fields in the non-relativistic limit.

In contrast, Marletto and Vedral’s field-based energy, as derived in Eq.~(\ref{eq:correct_field_energy}), is:

\begin{equation}
\mathcal{E}_{\text{field}} = \epsilon_0 \int \mathbf{E}_{\mathbf{s}} \cdot \mathbf{E}_{\mathbf{c}} \, d^3 \mathbf{r} + \frac{1}{\mu_0} \int \mathbf{B}_0 \cdot \mathbf{B}_{\mathbf{c}} \, d^3 \mathbf{r}.
\end{equation}
For static fields, this reduces to:

\begin{equation}
\mathcal{E}_{\text{field}} = q \phi_{\mathbf{s}}(\mathbf{r}_{\mathbf{c}}) + q \mathbf{v} \cdot \mathbf{A}_{\mathbf{s}}(\mathbf{r}_{\mathbf{c}}), \label{SF}
\end{equation}
as shown by integrating the electric and magnetic field overlaps. However, for time-dependent fields, the magnetic term requires correction due to Ampère’s law in the Coulomb gauge:

\begin{equation}
\nabla^2 \mathbf{A}_{\mathbf{c}} - \mu_0 \epsilon_0 \frac{\partial^2 \mathbf{A}_{\mathbf{c}}}{\partial t^2} = -\mu_0 \mathbf{j}_{\mathbf{c}},
\end{equation}
where \(\mathbf{A}_{\mathbf{c}}\) is the charge’s vector potential, and \(\mathbf{j}_{\mathbf{c}} = q \mathbf{v} \delta^3(\mathbf{r} - \mathbf{r}_{\mathbf{c}})\). The magnetic field integral becomes:

\begin{equation}
\frac{1}{\mu_0} \int \mathbf{B}_0 \cdot \mathbf{B}_{\mathbf{c}} \, d^3 \mathbf{r} = q \mathbf{v} \cdot \mathbf{A}_{\mathbf{s}} - \epsilon_0 q \int \frac{1}{4 \pi |\mathbf{r} - \mathbf{r}_{\mathbf{c}}|} \frac{\partial^2 \mathbf{A}_{\mathbf{s}}}{\partial t^2} \, d^3 \mathbf{r},
\end{equation}
introducing a time-dependent correction that breaks the equality \(\mathcal{E}_{\text{field}} = q \phi_{\mathbf{s}} + q \mathbf{v} \cdot \mathbf{A}_{\mathbf{s}}\) for general dynamic fields. Thus, their field-based energy (\ref{SF}) holds only approximately for static or slowly varying fields.

Comparing the energies, even for static fields where the equality holds:

\begin{equation}
\mathcal{E}_{\text{QED}} = q \phi_{\mathbf{s}} - q \mathbf{v} \cdot \mathbf{A}_{\mathbf{s}}, \quad \mathcal{E}_{\text{field}} = q \phi_{\mathbf{s}} + q \mathbf{v} \cdot \mathbf{A}_{\mathbf{s}},
\end{equation}
the magnetic term has opposite signs: \(-q \mathbf{v} \cdot \mathbf{A}_{\mathbf{s}}\) in QED versus \(+q \mathbf{v} \cdot \mathbf{A}_{\mathbf{s}}\) in their model. 
In the AB setup, where \(\phi_{\mathbf{s}} = 0\) (neutral solenoid):

\begin{equation}
\mathcal{E}_{\text{QED}} = -q \mathbf{v} \cdot \mathbf{A}_{\mathbf{s}}, \quad \mathcal{E}_{\text{field}} = q \mathbf{v} \cdot \mathbf{A}_{\mathbf{s}},
\end{equation}
highlighting the sign error. 

To sum up, \(\mathcal{E}_{\text{field}}\) does not align with \(\mathcal{E}_{\text{QED}}\), even in the Coulomb gauge, and fails for time-dependent fields. This invalidates the authors' claim that \(\mathcal{E}_{\text{field}}\) universally represents the local interaction driving the AB phase.

\subsubsection{Physical Critique of Local EM Field Mediation}

Marletto and Vedral’s claim that \(\mathcal{E}_{\text{field}}\) reflects a local EM field interaction mediating the AB phase is also  problematic. In the AB effect, the charged particle travels in a region where the EM fields vanish (\(\mathbf{B} = \nabla \times \mathbf{A} = 0\), \(\mathbf{E} = 0\) for \(r > a\)), but the vector potential \(\mathbf{A}\) is non-zero. The field-based energy \(\mathcal{E}_{\text{field}}\) involves \(\mathbf{B}_0\), which exists only inside the solenoid (\(r < a\)), not at the charge’s position (\(r > a\)). Thus, it cannot represent a local field interaction at the charge’s location. 

\subsection{Gauge Dependence of the AB Phase for Non-Closed Paths}

Marletto and Vedral also claim that the AB phase is gauge-independent for non-closed paths. 
However, this claim results from their incorrect assumption that the field-based interaction energy, \(\mathcal{E}_{\text{field}} = \frac{1}{\mu_0} \int \mathbf{B}_0 \cdot \mathbf{B}_{\mathbf{c}} d^3\mathbf{r}\), is valid in all gauges. Since \(\mathcal{E}_{\text{field}}\) is gauge-invariant (due to the gauge invariance of \(\mathbf{B}_0 = \nabla \times \mathbf{A}_{\mathbf{s}}\) and \(\mathbf{B}_{\mathbf{c}} = \nabla \times \mathbf{A}_{\mathbf{c}}\)), it produces a gauge-independent phase for non-closed paths:

\begin{equation}
\phi = \frac{\mathcal{E}_{\text{field}}}{\hbar} t,
\end{equation}
which contradicts standard QED predictions. This assumption stems from their belief that \(\mathcal{E}_{\text{field}}\) universally replaces the QED interaction energy -\(q \mathbf{v} \cdot \mathbf{A}_{\mathbf{s}}\), ignoring the sign mismatch and its gauge-specific derivation in the Coulomb gauge.

In standard QED, the AB phase is derived from the correct minimal coupling Hamiltonian, yielding the interaction Hamiltonian with the interaction energy -\(q \mathbf{v} \cdot \mathbf{A}_{\mathbf{s}}\). The phase for a path from \(\mathbf{r}_1\) to \(\mathbf{r}_2\) is:

\begin{equation}
\phi_{AB} = -\frac{q}{\hbar} \int_{\mathbf{r}_1}^{\mathbf{r}_2} \mathbf{A}_{\mathbf{s}} \cdot d\mathbf{l}.
\end{equation}
Under a gauge transformation \(\mathbf{A}_{\mathbf{s}}' = \mathbf{A}_{\mathbf{s}} + \nabla \chi\), the phase becomes:

\begin{equation}
\phi'_{AB} = -\frac{q}{\hbar} \int_{\mathbf{r}_1}^{\mathbf{r}_2} \mathbf{A}_{\mathbf{s}}' \cdot d\mathbf{l} = \phi_{AB} - \frac{q}{\hbar} \left[ \chi(\mathbf{r}_2) - \chi(\mathbf{r}_1) \right],
\end{equation}
demonstrating gauge dependence for non-closed paths, as \(\chi(\mathbf{r}_2) \neq \chi(\mathbf{r}_1)\) in general. This gauge dependence is a hallmark of the AB effect in QED for non-closed paths, only becoming gauge-invariant for closed paths where \(\oint \nabla \chi \cdot d\mathbf{l} = 0\). Marletto and Vedral’ model, by relying on a gauge-invariant \(\mathcal{E}_{\text{field}}\), fails to reproduce this gauge dependence, indicating a deviation from standard QED predictions.

\subsection{Entanglement is Not the Primary Cause of the AB Phase}
\label{sec:entanglement}

Marletto and Vedral's central claim is that the AB phase arises from local entanglement between the charged particle and the photon field. While their QED model does indeed produce such entanglement, we demonstrate that this entanglement is merely a consequence of the interaction formalism rather than the physical mechanism responsible for the AB phase. The phase is determined by the vector potential $\mathbf{A}$, with entanglement playing no causal role.

\subsubsection{The Origin of the Phase}

The interaction Hamiltonian in their model (after sign correction),
\begin{equation}
H_{\text{int}} = -\frac{q}{m} \mathbf{p} \cdot \mathbf{A}(\mathbf{r}_c) q_z^{(C)} - \int d^3 \mathbf{x} \, \mathbf{j}(\mathbf{x} - \mathbf{r}_{\mathbf{s}}) \cdot \mathbf{A}(\mathbf{x}) q_z^{(S)},
\end{equation}
couples the charged particle's momentum $\mathbf{p}$ to the quantized vector potential $\mathbf{A}$. For a particle in a superposition of left ($|L\rangle_C$) and right ($|R\rangle_C$) paths around the solenoid, this interaction generates path-dependent phases:
\begin{equation}
\phi_L = -\frac{q}{\hbar} \mathbf{v} \cdot \mathbf{A}(\mathbf{r}_{\mathbf{L}}) \tau, \quad \phi_R = -\frac{q}{\hbar} \mathbf{v} \cdot \mathbf{A}(\mathbf{r}_{\mathbf{R}}) \tau,
\end{equation}
where $\tau$ is the interaction time. The phase difference,
\begin{equation}
\Delta \phi = \phi_R - \phi_L = -\frac{q}{\hbar} \oint \mathbf{A} \cdot d\mathbf{l},
\end{equation}
matches the standard AB phase and depends \textit{only} on the $\mathbf{A}$-field. 

\subsubsection{Incidental Nature of Entanglement}

The system's post-interaction state is:
\begin{equation}
|\psi\rangle = \frac{1}{\sqrt{2}} \left( e^{i \phi_L} |L\rangle_C |\chi_L\rangle_F + e^{i \phi_R} |R\rangle_C |\chi_R\rangle_F \right) |1\rangle_S,
\end{equation}
where $|\chi_L\rangle_F$ and $|\chi_R\rangle_F$ are photon states associated with each path. Tracing out the photon field yields the reduced density matrix for the charge:
\begin{equation}
\rho_C = \text{Tr}_F \left( |\psi\rangle \langle \psi| \right) = \frac{1}{2} \left( |L\rangle \langle L| + |R\rangle \langle R| + e^{i \Delta \phi} \langle \chi_L|\chi_R\rangle_F |L\rangle \langle R| + \text{h.c.} \right).
\end{equation}
Crucially, in the AB regime, the photon field perturbation is negligible ($|\chi_L\rangle_F \approx |\chi_R\rangle_F$), so $\langle \chi_L|\chi_R\rangle_F \approx 1$. Then the off-diagonal terms ($|L\rangle \langle R|$) retain the phase $\Delta \phi$ \textit{independent} of the photon overlap. 
Thus, while entanglement exists, it does not influence the observable phase.

\subsubsection{Semiclassical Consistency}

The AB phase can be derived \textit{without} invoking entanglement:
In the path integral formulation, the phase arises from $\exp \left( -\frac{iq}{\hbar} \int \mathbf{A} \cdot d\mathbf{l} \right)$, with no reference to photon states, and the semiclassical treatment \cite{aharonov1959} uses only the classical $\mathbf{A}$-field.
This reinforces that entanglement in Marletto and Vedral's model is not a physical requirement. 

To sum up, while Marletto and Vedral's model formally introduces entanglement between the charge and photon field, this entanglement:
(1) Does not determine the AB phase (which is fixed by $\mathbf{A}$); (2) Has negligible effect on observables ($\langle \chi_L|\chi_R\rangle_F \approx 1$); and (3) Is absent in simpler derivations of the effect. 
Thus, the AB phase remains a manifestation of the vector potential's role, not quantum correlations with the photon field.

\section{Conclusion}

Our analysis of Marletto and Vedral’s model reveals critical flaws: an incorrect \(\frac{1}{2}\) prefactor in their field-based energy, halving the interaction strength; a sign error yielding \( +q \mathbf{v} \cdot \mathbf{A}_{\mathbf{s}} \) instead of QED’s \( -q \mathbf{v} \cdot \mathbf{A}_{\mathbf{s}} \), even in the Coulomb gauge; a false claim of gauge independence for non-closed paths, inconsistent with QED’s gauge-dependent phase; and an unsubstantiated role for entanglement, which is incidental rather than causal. The field-based energy’s equivalence to \( q \mathbf{v} \cdot \mathbf{A}_{\mathbf{s}} \) holds only approximately in the Coulomb gauge under static conditions, failing for time-dependent fields or other gauges. These errors misrepresent the AB effect’s mechanism. This Comment corrects these fundamental flaws, reinforcing the conventional explanation of the AB effect in the semi-classical picture, where the AB phase is driven by the local coupling to the vector potential \(\mathbf{A}\).

\bibliographystyle{plain}

\end{document}